\documentstyle[12pt,graphics]{article}

\title{Modelling the flyby anomalies using a modification of inertia.}

\author{M.E.~McCulloch \footnote{School of Physics, University of Exeter, Devon, UK (M.E.McCulloch@exeter.ac.uk)}}

\begin{document}

\maketitle

\newpage

\section*{ABSTRACT}

The flyby anomalies are unexplained velocity jumps of 3.9, -4.6, 13.5, -2, 1.8 
and 0.02 mm/s observed near closest approach during the Earth flybys of six 
spacecraft. These flybys are modelled here using a theory that assumes that 
inertia is due to a form of Unruh radiation, and varies with acceleration due 
to a Hubble-scale Casimir effect. Considering the acceleration of the craft 
relative to every particle of the rotating Earth, the theory predicts that 
there is a slight reduction in inertial mass with increasing latitude for 
an unbound craft, since near the pole it sees a lower average relative 
acceleration. Applying this theory to the in- and out-bound flyby paths, 
with conservation of momentum, the predicted anomalies were 2.9, -0.9, 20.1, 
0.9, 3.2 and -1.3 mm/s. Three of the flyby anomalies were reproduced within 
error bars, and the theory explains their recently-observed dependence on 
the latitude difference between their incident and exit trajectories. The 
errors for the other three flybys were between 1 and 3 mm/s.

\vspace{2cm}

KEYWORDS: solar system: general, cosmology: theory, gravitation, celestial mechanics.

\newpage

\section{INTRODUCTION}

During six Earth gravity assist flybys, significant anomalous velocity 
increases of a few mm/s were observed (Antreasian and Guinn,~1998 and 
Anderson~$et~al.$,~2008) using both doppler frequency data and ranging 
methods. These are known as the flyby anomalies, and, so far, no 
explanations have been found that can account for them. Explanations 
considered and rejected to date have included: anomalous thruster 
activity, computer software glitches, troposphere and ionosphere effects 
and others (see Antreasian and Guinn,~1998 and Lammerz\"ahl~$et~al.$~(2006).

Anderson~$et~al.$~(2008) analysed the available data, some of which are 
summarised in table 1 (columns 1-4), and intriguingly managed to show 
that the six velocity anomalies observed so far (dv) fitted a formula, 
given, with slightly modified notation, by
\begin{equation}
dv = 3.099 \times 10^{-6} \times v_{\infty} \times (cos {\phi}_1 - cos {\phi}_2)
\end{equation}
where $v_{\infty}$ is the hyperbolic excess velocity, ${\phi}_1$ is the 
incident angle of the trajectory and ${\phi}_2$ is the exit angle. Equation 
(1) shows that the anomalous velocity gain $dv$ depends on the difference 
between the incident latitude and the latitude of the exit trajectory. 
For example, the NEAR (Near Earth Asteroid Rendezvous) probe approached 
at low latitude and left on a polar trajectory and its velocity jump was 
large, whereas Messenger approached and left on an equatorial trajectory 
and only a very small jump was seen. This led Anderson~$et~al.$~(2008)
to make the interesting suggestion that the cause may be somehow related 
to the Earth's rotation, although they did not suggest a cause. A possible 
cause is suggested in this paper.

McCulloch~(2007) proposed a model in which the inertial mass reduces slightly 
as the acceleration decreases: a modification of inertia due to a Hubble-scale 
Casimir effect (hereafter: MiHsC). This is interesting, because if we take 
the unusual step of summing all the accelerations seen by NEAR on its flyby, 
then on its equatorial approach it would see high accelerations as the masses 
comprising the planet rotate towards and away from it, and many of the 
acceleration vectors would point at the craft, but on its polar exit 
trajectory, NEAR would see much less acceleration since the Earth's 
acceleration vectors, pointing at the spin axis, would not point at the 
craft. Therefore, MiHsC predicts a lower post-flyby inertial mass for NEAR, 
which, through conservation of momentum, implies an increase in its speed. 
In this paper it is shown that the increase in speed predicted by MiHsC 
agrees quite closely with the observed flyby anomalies.

\section{METHOD AND RESULTS}

Haisch~$et~al.$~(1994) suggested that inertial mass could be caused by a 
form of Unruh radiation. Milgrom~(1994,1999) suggested that there could be 
an abrupt break in this effect for very low accelerations since the Unruh 
wavelengths would then exceed the Hubble distance, and the loss of inertia 
for low acceleration would be similar to the behaviour of his empirical 
MOND (Modified Newtonian Dynamics) scheme. The model of McCulloch~(2007) 
builds on this suggestion, but uses a Hubble-scale Casimir effect instead 
so that inertia diminishes linearly as acceleration reduces, since fewer 
wavelengths fit within the Hubble diameter. This is a more gradual process 
than the abrupt break of Milgrom. This model could be called Modified 
Inertia due to a Hubble-scale Casimir effect (or MiHsC). 
In MiHsC the equivalence principle ($m_i$=m$_g$) is changed slightly to
\begin{equation}
m_I = m_g \left( 1- \frac{\beta \pi^2 c^2}{a \Theta} \right).  \label{eq:mi}
\end{equation}
where $m_I$ is the modified inertial mass, $m_g$ is the gravitational mass 
of the spacecraft, $\beta=0.2$ (from the empirically-derived Wien's 
constant), $c$ is the speed of light, $\Theta$ is twice the Hubble distance 
$2c/H$, and $a$ is the acceleration of the craft relative to the matter in 
its local environment. 
In McCulloch~(2007) this was simplified to be the acceleration of the 
Pioneer craft relative to the Sun's centre of mass (Ignatiev~2007, in 
his version of modified inertia, uses an acceleration relative to the 
galactic centre). At most terrestrial values of acceleration the difference 
from standard physics is small, but this model predicts the Pioneer 
anomaly correctly beyond 10 au from the Sun with no adjustable 
parameters (McCulloch,~2007). 

  However, the model also predicted an anomaly within 10au of the Sun, 
  (when the Pioneer craft were in bound orbits, and no anomaly was observed). 
  The model is also not needed to explain the orbits of the planets, and 
  its variation of inertial mass disagrees with precise Earth-bound tests 
  of the equivalence principle undertaken by, for example, 
  Carusotto~$et~al.$~(1992) and Schlamminger~$et~al.$~(2008).
  These results suggest that, if this model is correct, it only applies to
  unbound orbits. The reason for this is unknown. However, the work of 
  Price (2005) is interesting in this respect because he showed that a bound 
  system does not follow the cosmological expansion, whereas an unbound system does.

To analyse the trajectories of the flyby craft (which are not bound to the 
Earth) we first assume conservation of momentum so that
\begin{equation}
m_{1e}v_{1e}+m_1 v_1 = m_{2e}v_{2e} + m_2 v_2
\end{equation}
where the terms are: the initial momentum of the Earth, the initial 
momentum of the craft, and the final momenta. We now replace the 
inertial masses of the unbound craft $m_1$ and $m_2$ with the modified 
inertia of McCulloch~(2007) (equation 2) so that
\begin{equation}
m_{1e}v_{1e} + m_g \left(1-\frac{\beta {\pi}^2 c^2}{a_1 \Theta} \right) v_1 = m_{2e}v_{2e} + m_g \left(1-\frac{\beta {\pi}^2 c^2}{a_2 \Theta} \right) v_2
\end{equation}
where $m_g$ is the gravitational mass of the craft, or the uncorrected 
inertial mass. Some algebra implies that
\begin{equation}
v_2-v_1 = dv = \frac{m_e}{m_g} (v_{1e}-v_{2e}) + \frac{\beta \pi^2 c^2}{\Theta} \left( \frac{v_2}{a_2} - \frac{v_1}{a_1} \right)
\end{equation}
The first term on the right hand side is well known. So we now look at the 
new velocity change due to modified inertia represented by the second term 
and call it $dv'$. 
\begin{equation}
dv' = \frac{\beta \pi^2 c^2}{\Theta} \left( \frac{v_2}{a_2} - \frac{v_1}{a_1} \right)
\end{equation}
We take the incoming and outgoing craft at a radius where the standard 
gravitational acceleration is equivalent, so by standard physics $a_1=a_2$, 
but take the new step of assuming that the accelerations of the craft 
relative to each part of the Earth also contributes to $a_1$ and $a_2$. 
  To picture these accelerations one could imagine a line connecting 
  the craft with every mass in the Earth and measure the acceleration of 
  the length of each line to determine the inertial mass to use for each 
  gravitational interaction.
It can be shown that the average acceleration of particles in the x-direction 
(the assumed direction of the craft) within the solid Earth 
$a=0.07v_e^2/R$ (see the Appendix and Figure 1 for a derivation) 
where $v_e$ is the rotational velocity at the surface equator, and $R$ is 
the Earth's radius. The component of the acceleration seen by a flyby craft 
at a latitude $\phi$ (Fig.~1) is therefore 
$a=(0.07v_e^2 / R ) \times cos \phi$. 
We now use these accelerations in equation~(6) and get
\begin{equation}
dv' = \frac{\beta{\pi}^2 R c^2}{0.07 \times v_e^2 \Theta} \times \left( \frac{v_2 cos{\phi}_1 - v_1 cos{\phi}_2}{cos{\phi_1} cos{\phi}_2} \right)
\end{equation}
Substituting values as follows $R=6371~km$, $c=3 \times 10^8~m/s$, 
$v_e=465~m/s$ and the Hubble diameter $\Theta=2c/H=2.7 \times 10^{26}~m$, 
the same value used by McCulloch~(2007) to reproduce the Pioneer anomaly
\begin{equation}
dv' = 2.8 \times 10^{-7} \times \left( \frac{v_2 cos{\phi}_1 - v_1 cos{\phi}_2}{cos{\phi}_1cos{\phi}_2} \right)
\end{equation}
The derived equation (8) is similar to the empirical equation derived by 
Anderson~$et~al.$~(2008) (equation 1) from their data, especially the 
dependence on the difference in the cosine of the incident and outgoing 
latitude (the denominator in the brackets has little effect on most of 
the flybys being a little less than 1 for most of them, and 0.3 for NEAR).

Equation (8) was used to predict the flyby anomalies and the results are 
shown in table~1, column 5 and in Figure 2. The errors were calculated 
assuming a 9 $\%$ error in the Hubble constant (Freedman~$et~al.$~2001), 
and an error caused by the assumption in the Appendix that it is the 
x-component of the acceleration that matters: assuming that the craft is
at an infinite distance from the Earth. It is, more properly, the component 
of acceleration pointing along the aforementioned lines between the craft
and each point in the Earth that matters. The error from this source, was 
calculated by assuming a distance from the Earth of 36,000~km (roughly the
distance from which the post-encounter data was available). The average 
error in the acceleration vector's orientation by taking only the 
x-component is then about 3$^o$. To calculate the errors, the $\phi_1$ 
and $\phi_2$ in equation 8 were each alterred by 3$^o$. The resulting 
variations in the predicted $dv$ were 0.5, 0.5, 6.4, 0.2, 0.5 and -0.04. 
These were added to the errors due to the Hubble constant and a ten 
percent error in the assumed linear vertical density profile of the Earth 
(see the Appendix). The resulting error bars are shown in Figure 2.

In Figure 2 the flyby passes are shown one by one along the x-axis, the 
diamonds show the observed velocity jump (The observational error was 
assumed to be 0.1~mm/s) and the pluses show the predicted jump using 
equation (8), with error bars. 

The predictions agree with the observations for the Galileo-I, NEAR and 
Rosetta flybys. In the other cases there are differences. For Galileo-II 
the difference is 3 mm/s, although it is possible that the observed jump 
in this case may have a larger error since this flyby grazed the atmosphere
and had to be corrected for atmospheric drag.

Nevertheless, the dependence on change of latitude seen by Anderson 
$et~al.$ (2008) is reproduced. The predicted values are in proportion 
to those observed, and the correlation between the six observed and 
predicted values is 0.94 (although this is not significant at the 
5$\%$ level: the p value is 30). It would be useful to have 
a larger set of observations to assess the theory.

  Flybys of other planets or moons would provide an interesting test of
  this model, since the planet's radius $R$ and especially its equatorial
  velocity $v_e$ should strongly effect the size of the anomaly (see Eq.7).
  A large slowly-rotating planet, or even a galaxy, could show a strong,
  and more detectable, velocity boost for polar exit trajectories. The 
  recent (February 2007) Rosetta flyby of Mars could be useful, since
  the values of $R$ and $v_e$ for Mars imply that its flyby anomaly should 
  be approximately double that of the Earth.

The predictions of equation (8) are not as close as the predictions of 
Anderson~$et~al.$'s suggestive equation (1), although their empirical formula 
was, of course, fitted to the data, whereas equation (8) was derived from a 
theory.

\section{CONCLUSION}

Six well-observed Earth flybys with unexplained velocity anomalies were 
modelled using a theory that assumes that inertia is due to a form of 
Unruh radiation, and varies with acceleration due to a Hubble-scale 
Casimir effect. 

The theory reproduces three of the observed flyby anomalies within error 
bars, without the need for adjustable parameters, and explains the 
recently-observed dependence of the anomalies on the latitude difference 
between the incident and exit trajectories. The errors for the other 
flybys were 1, 2 and 3 mm/s.

It should be stressed that the suggested model only seems to apply to unbound
trajectories and the reason for this is unknown. Also, the definition of 
acceleration used here is different to that used in Milgrom~(1994,~1999) 
and Ignatiev~(2007). The method uses the same physics as McCulloch~(2007)
(eg: Eq. 2), but differs in including more detailed information about 
accelerations. The results of McCulloch~(2007) would be largely 
unchanged using this enhanced scheme since the Pioneer craft maintained 
a low latitude with respect to the Sun (Parthasarathy and King, 1991).

\section*{ACKNOWLEDGEMENTS}

Many thanks to D.Price, C.Smith, B.Kim, R.McCulloch and an anonymous reviewer 
for useful discussions or comments.

\section*{APPENDIX}

The magnitude of the acceleration of a point a distance $r$ from the 
rotational axis of a uniform sphere is $a=v^2/r$. Expressing this in 
terms of the equatorial rotational velocity of the Earth $v_e$ we get 
$a={v_e}^2r / R^2$ where $R$ is the Earth's equatorial radius. 
To calculate the average acceleration seen by the flyby craft, we assume 
they are far enough from the planet along the x axis, and therefore see 
the x-coordinate of $a$ which, in terms of the longitude $\lambda$ and 
angle from the north pole $\theta$ (see Figure 1) is
\begin{equation}
a_x=\frac{{v_e}^2 r}{R^2} sin \lambda ~ sin \theta
\end{equation}
we use $r=c sin \theta$, where $c$ is the distance from the centre
of the Earth
\begin{equation}
a_x=\frac{{v_e}^2 c sin \theta}{R^2} sin \lambda ~ sin \theta
\end{equation}
We integrate this over the sphere to find the average acceleration of all 
the mass in the Earth. Since the density of the core of the Earth is 
12.8-13.1 $g/cm^3$ and that of the crust is only 2.2-2.9 $g/cm^3$ 
(Dzievonski and Anderson,~1984) we need to weight the integral higher 
towards the centre. For simplicity we assume a linear increase of density 
with depth which can be modelled as 
$\rho=(R- \alpha c)/R$ 
where $\alpha=0.974$. In the integral $\phi$ is the longitude and $\theta$ 
is the angle from the north pole. A volume element is therefore 
$dc.csin \theta d \phi. c d \theta$
\begin{equation}
\overline{a_x}=\frac{1}{4/3 \pi R^3} \times \int_{c=0}^{R} 2 \int_{\lambda=0}^\pi \int_{\theta=0}^\pi \frac{v_e^2}{R^2} c^3 \frac{R-\alpha c}{R} (sin \theta)^3 ~ sin \lambda ~ dc d\lambda d\theta
\end{equation}
The result is
\begin{equation}
\overline{a_x}=\frac{4v_e^2}{\pi R} \left( \frac{1}{4} - \frac{\alpha}{5} \right) \sim 0.07 \times \frac{v_e^2}{R}
\end{equation}

\section*{REFERENCES}

Anderson,~J.D.,~J.K.~Campbell, J.E.~Ekelund, J.~Ellis and J.F.~Jordan, 2008.
$Phys.~Rev.~Lett.$, 100, 091102.

Antreasian,~P.G. and J.R.~Guinn,~1998. Investigations into the unexpected
delta-v increases during the Earth gravity assists of Galileo and NEAR.
Paper no. 98-4287 presented at the AIAA/AAS Astrodynamics Specialist Conf.
and Exhibition, Boston, 1998.

Carusotto,~S., V.~Cavasinni and A.~Mordacci, 1992. Test of G-universality
with a Galileo type experiment. $Phys.$ $Rev.$ $Lett.$, 69, 12, 722-1725.

Dzievonski,~A.M. and D.L.~Anderson, 1984. Seismic tomography of the Earth's
interior. $Am.~Sci.$, 72(5), 483-494.

Freedman,W.L.~2001. Final results from the Hubble space telescope key project
to measure the Hubble constant. $ApJ$, 553, 47-72.

Haisch,~B., Rueda,~A. and Puthoff,~H.E., 1994. Inertia as a zero-point field 
Lorentz force. $Phys.~Rev.~A$,~49,~678.

Ignatiev,~A.Y.,~2007. Is violation of Newton's second law possible.
$Phys.$ $Rev.$ $Lett.$, 98, 101101. arxiv:gr-qc/0612159.

Lammerz\"ahl,~C, O~Preuss, H.~Dittus,~2006. arxiv:gr-qc/0604052.

McCulloch,~M.E.,~2007. Modelling the Pioneer anomaly as modified inertia.
$MNRAS$, 376, 338-342. arxiv:astro-ph/0612599.

McGaugh,~S.S.,~2007. Modified Newtonian Dynamics close to home (reply).
$Science$, Letters, 318, 568-570.

Milgrom,~M.,~1994. Ann.Phys., 229, 384.

Milgrom,~M.,~1999. The Modified Dynamics as a vacuum effect. 
$Phys.~Lett.~A$, 253, 273.

Parthasarathy, R and J.H.King, 1991. Trajectories of Inner and Outer 
Heliospheric Spacecraft Predicted Through 1999. 
http://cohoweb.gsfc.nasa.gov/

Price,~R.H.,~2005. In an expanding universe, what doesn't expand? arxiv:gr-qc/0508052

Schlamminger, S., K.Y.~Choi, T.A. Wagner, J.H. Gundlach and E.G.~Adelberger, 2008.
Test of the equivalence principle using a rotating torsion balance.
$Phys.$ $Rev.$ $Lett$, 100, 4, 041101.

\newpage

\begin{center}
FLYBYS: OBSERVED AND PREDICTED
\begin{table}[htp]
\begin{center}
\begin{tabular}{l|c|c|c|c}                                           \hline
Mission    & Flyby speeds & Latitudes   & Observed dV           & Predicted      \\
           &    (km/s)    &  (deg)      &  (mm/s)              & (mm/s)         \\ \hline\hline
Galileo-I  &    31,35     & -12.5,-34.2 &\bf{3.92} $\pm 0.08$  & \bf{ 2.9} $\pm 0.6$ \\ \hline
Galileo-II &    34.5,38.5 & -34.3,-4.9  &\bf{-4.6} $\pm 0.08$  & \bf{-0.9} $\pm 0.2$ \\ \hline
NEAR       &    36.5,34   & -20.8,-72   &\bf{13.46} $\pm 0.13$ & \bf{20.1} $\pm 4.0$ \\ \hline
Cassini    &    35,39     & -12.9,-5    &\bf{-2}               &  \bf{0.9} $\pm 0.2$ \\ \hline
Rosetta    &    31,35     & -2.8,-34.3  &\bf{1.8}  $\pm 0.05$  &  \bf{3.2} $\pm 0.6$ \\ \hline
Messenger  &    29,25     & 31.4,-31.9  &\bf{0.02}  $\pm 0.05$ & \bf{-1.3} $\pm 0.2$ \\ \hline
\end{tabular}
\end{center}
\caption{Flybys: observed and predicted. The mission name (column 1), the initial and 
final flyby speed (column 2), the initial and final latitudes (3), the observed 
anomalous velocity jumps (4) (from Anderson~$et~al.$~2008). The error bars are 
also shown, where known. The predictions made in this paper are shown in column 5.}
\end{table}
\end{center}

\newpage

\begin{center}
\resizebox{350pt}{350pt}{\includegraphics{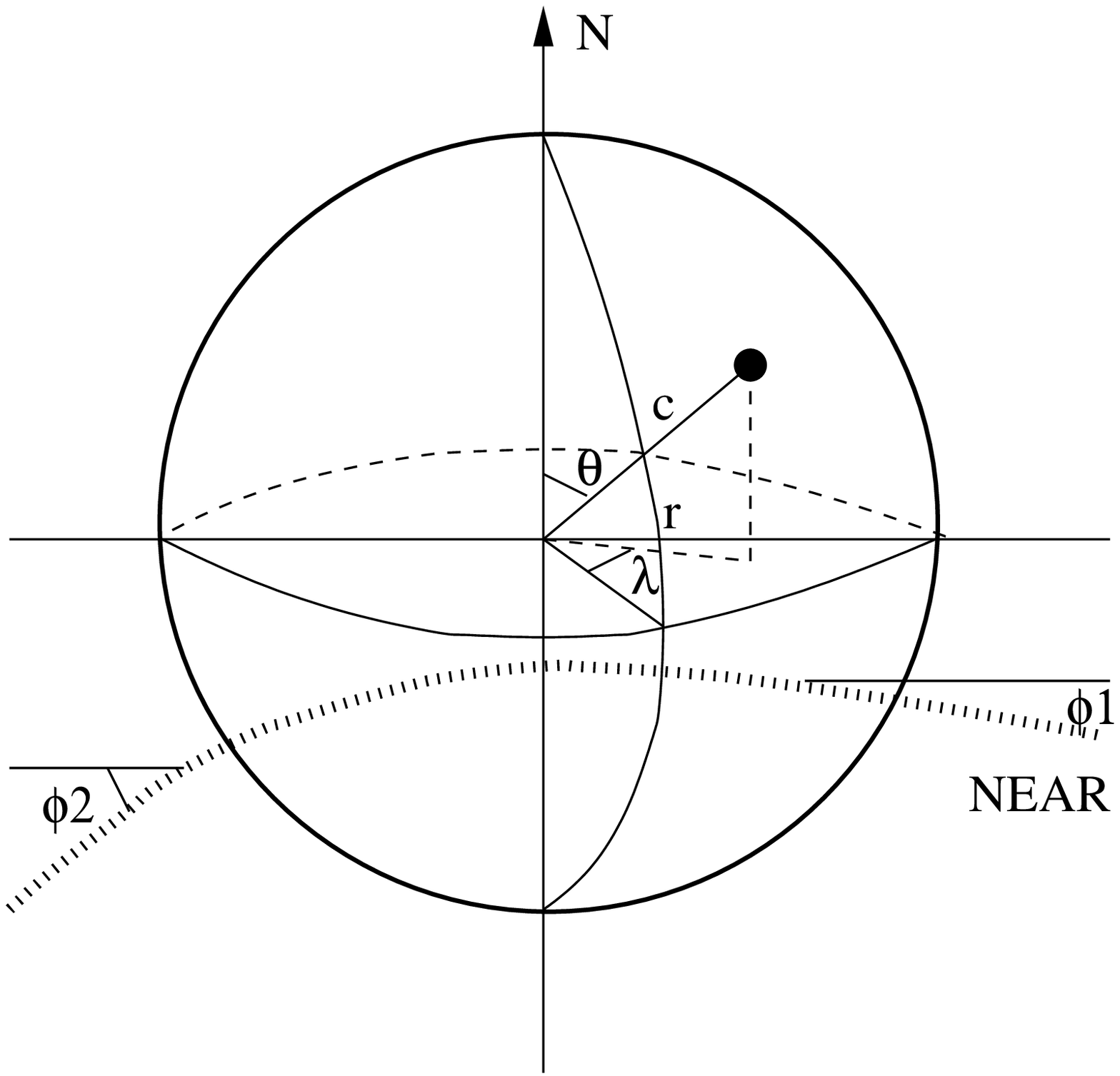}}
\end{center}

Figure 1. Schematic showing the NEAR flyby. This was incident at $\phi_1=-21^o$ latitude 
and left the Earth at $\phi_2=-72^o$ latitude. Also shown are some of the parameters 
used in the text and Appendix.

\newpage

\begin{center}
\resizebox{350pt}{350pt}{\includegraphics{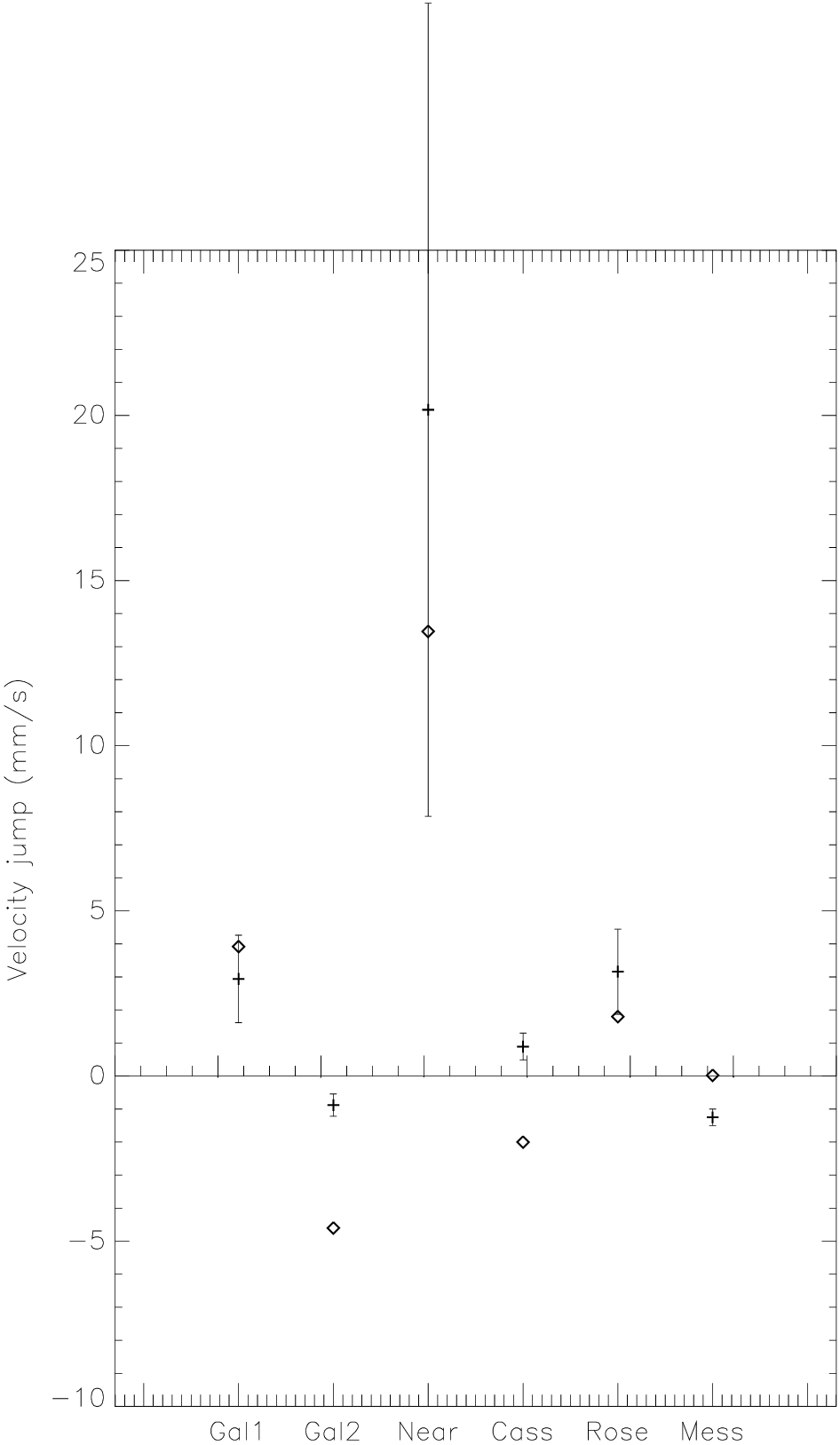}}
\end{center}

Figure 2. Observed (diamonds) and predicted (pluses) velocity jumps for all six 
flybys in mm/s with error bars shown for the predictions. The predictions agree
for the Galileo-I, NEAR and Rosetta flybys, and the differences are about 1~mm/s 
for Messenger, 2~mm/s for Cassini and 3~mm/s for Galileo-II.

\end{document}